\documentstyle[preprint,aps,bezier]{revtex}

\begin{document}

\title{Lagrangian Becchi-Rouet-Stora-Tyutin treatment of collective
coordinates}

\author{
Juan P.\ Garrahan,\thanks{Electronic address:
garrahan@tandar.cnea.edu.ar.} Mart\'{\i}n Kruczenski, and Daniel R.\ Bes}

\address{Departamento de F\'{\i}sica, Comisi\'on Nacional de Energ\'{\i}a
At\'omica, Av.\ Libertador 8250, 1429 Buenos Aires, Argentina.}

\date{November 3, 1995}
\preprint{TAN-FNT-95/013}

\maketitle

\begin{abstract}
The Becchi-Rouet-Stora-Tyutin (BRST) treatment for the quantization of
collective coordinates is considered in the Lagrangian formalism.  The
motion of a particle in a Riemannian manifold is studied in the case
when the classical solutions break a non-abelian global invariance of
the action.  Collective coordinates are introduced, and the resulting
gauge theory is quantized in the BRST antifield formalism.  The
partition function is computed perturbatively to two-loops, and it is
shown that the results are independent of gauge-fixing parameters.

\end{abstract}

\pacs{11.10.Lm}

\section{Introduction}

The correct quantization of collective coordinates has been a central
issue in the study of soliton models. For example, in the Skyrme model
\cite{Sk61} baryons are solitons of an $SU(2)_L \times SU(2)_R/SU(2)_V$
quiral Lagrangian and internal degrees of freedom (spin and isospin)
are described as collective excitations of the soliton \cite{Wi79}.
These collective excitations typically appear as zero energy
fluctuations around a classical solution which breaks certain invariance
of the action. To deal with the zero modes
the degrees of freedom associated with the collective motion should be
isolated.  As this is difficult to do in general, collective
coordinates can be introduced explicitly as {\it additional} degrees of
freedom. This procedure has become known as the collective coordinate
method \cite{GJ75,GS75,CL75}.

The general way to introduce collective variables is as parameters of
transformations of the original fields \cite{HK75}. The transformed
theory, which depends upon the overcomplete set of original fields plus
collective variables, is a gauge theory.

The most rigorous treatment of gauge theories is the
Becchi-Rouet-Stora-Tyutin (BRST) quantization procedure
\cite{BFV75,HT92}. The application of the BRST method to the
quantization of collective coordinates in finite systems has been
developed in \cite{KBC88,BCK89,KBC90,BK90}. The BRST quantization of
collective coordinates or fields in the case of field theory was
presented in \cite{AD90}.  Simple solitonic models have also been
studied using these framework \cite{ABS92,GKSBS95}.

In finite systems and soliton models, collective coordinates are
introduced as parameters of the symmetry transformations which are
broken (or partially broken) by the classical solutions.\footnote{The
case in which the transformations are neither restricted to form a
group nor to be symmetries of the action is discussed in \cite{AD90}.}
The resulting gauge theory is quantized
following the usual canonical BRST procedure. Lagrange multipliers and
ghosts are introduced for each gauge freedom. An hermitian, nilpotent
BRST charge is constructed, and physical states and operators are
defined as those invariant under the global BRST symmetry. The gauge
theory is thus replaced by a gauge-fixed BRST theory, in which the
collective variables become physical, and the zero modes are cancelled
by the Lagrange multipliers and the ghosts.

The BRST quantization can be formulated as a Lagrangian path integral
by means of the antifield formalism \cite{BV81,GPS94}. Given a gauge
theory with its corresponding ghosts, antifields are introduced,
and an antibracket structure is defined. The gauge-fixed BRST action is
obtained from the solution of the so-called Master Equation. Everything
is formulated directly in the space of Lagrangian variables.

The purpose of the present paper is to apply the BRST treatment of
collective coordinates in the Lagrangian path integral framework to the
motion of a particle in a Riemannian manifold. This model includes the
Skyrme and the $O(3)$ models, as well as very simple cases which can be
used to verify the calculations \cite{GKB95}.

Although this model could be treated with the usual Faddev-Popov
techniques, the antifield method should be necesary to treat
for example supersymmetric solitons when the supersymmetric
algebra closes on-shell (i.e. up to terms proportional to equations 
of motion).  

The outline of this paper is as follows. In Sec.\ref{SM} we describe
the model and the group $G$ of transformations that leave the action
invariant. The classical solution to the equations of motion, as shown
in Sec.\ref{SSCC}, is only invariant under a subgroup $H$ of $G$,
giving rise to zero modes. Collective coordinates have to be introduced
to restore the broken symmetry. This yields a gauge invariant action.

In Sec.\ref{AF} the antifield formalism is applied to the quantization
of the gauge invariant action. A BRST gauge-fixed Lagrangian action is
constructed, and the path integral for the quantization is set up. An
equivalent result can be obtained from the Hamiltonian approach, as
mentioned in Sec.\ref{HBQ}.

Sec.\ref{PT} discusses the perturbative computation of the partition
function. Expressions for the free propagators and interaction vertices
are given. Two-loops corrections to the intrinsic energy levels are
calculated (formally) with the aid of finite temperature techniques,
and a collective Hamiltonian to one loop is obtained. These results
are shown to be free of any dependence upon spurious parameters
introduced by the gauge-fixing procedure. 

One aspect not treated in the present paper is the ultraviolet
renormalization since it depends on the particular model to which the
formalism is applied.

Finally, our conclusions and outlook are given in Sec.\ref{SC}. The
Appendix contains definitions and relations used in calculating the
diagrams.

\section{The model} \label{SM}

Consider a particle moving in a Riemannian manifold $M$ with metric
$g_{st}$. The Euclidean action is taken to be:
\begin{equation}
S = \int d\tau \left( 
	\frac{1}{2} g_{st} \dot{q}^s \dot{q}^t + V[q] 
	\right) , \label{L}
\end{equation}
The coordinates $q^s$ parametrize the manifold and $V[q]$ is an arbitrary
potential. The dot $\dot{q}^s$ denotes derivatives respect to Euclidean 
time $\tau$. Summation is implicit over repeated indices $s,t,\ldots$,
summation over other indices are written explicitly.

A point $q^s$ in $M$ describes a configuration of the system. For
example in the Skyrme model $M$ is the (infinite dimensional) manifold
of functions (of winding number one) $q:S^3\rightarrow SU(2)$, where
$S^3$ is the compactified physical space and $SU(2)$ is parametrized by
the pions. In this case the variable $q$ is a scalar field
$\phi_a(\vec{x}), a=1,2,3$. In the following, $s$ stands for both $a$
and $\vec{x}$.  Sums in $s$ are then understood as sums over $a$ and
integrals over $\vec{x}$. With this conventions, our formalism can be
applied to soliton models as well as to simple quantum mechanical
models.
 
We assume that there exists an unbroken non-trivial, finite-dimensional,
 group $G$ of
symmetries of the action, constituted by isometries which also leave
the potential invariant:
\begin{eqnarray}
q^s &\rightarrow& R^s(q,\alpha_a) , \\
g_{st}(q) &=& \partial_s R^u
              \partial_t R^v g_{uv}(R(q,\alpha_a)) , \label{varg} \\
V[q] &=& V[R(q,\alpha_a)] , \label{varV}
\end{eqnarray}
where the $\alpha_a$'s parametrize the group $G$. By $\partial_s$ and 
$\partial_a$ we mean $\frac{\partial}{\partial q^s}$ and 
$\frac{\partial}{\partial \alpha_a}$. The parametrization is such that
$\alpha_a=0$ corresponds to the identity $(R^s(q,0)=q^s)$.

Infinitesimal variations, which generate the Lie algebra ${\bf g}$ of $G$ 
are defined as:
\begin{equation}
\delta_a q^s = \left. \partial_a R^s(q,\alpha) \right|_{\alpha=0} .
\end{equation}
Commuting two such infinitesimal variations we obtain:
\begin{equation}
\partial_a R^t \partial_b \partial_t R^s - 
	\partial_a \partial_t R^s \partial_b R^t =
	{C_{ab}}^c \partial_c R^s , \label{commR}
\end{equation}
where ${C_{ab}}^c$ are the structure constants of the group $G$.
 
In the next section a semiclassical expansion is performed around a
minimum of the potential $V$. The interesting case is that in which
there are several classical minima related by transformations of $G$
whereas quantum mechanically there is only a single vacuum because $G$
is unbroken (this is not the case of spontaneous symmetry breaking in
which there are many {\it quantum} vacua). Choosing a particular
classical minimum brings in infrared divergencies which spoil the
calculation (excitations tangent to the surface of minima have zero
restoration frequency). The well known cure is the introduction of
collective coordinates which eliminate the zero modes and at the same
time restore the symmetry which was broken at the classical level.

\section{Static solutions and collective coordinates} \label{SSCC}

The equations of motion obtained from (\ref{L}) are
\begin{equation}
g_{st} \ddot{q}^t + \frac{1}{2} \dot{q}^t \dot{q}^u \left( 
	\partial_u g_{st} + \partial_t g_{su} - \partial_s g_{tu} 
	\right) - \partial_s V = 0 .
\end{equation}
Static solutions satisfy the time-independent
equation
\begin{equation}
\left. \partial_s V \right|_{q = \bar{q}} = 0 . \label{min}
\end{equation}

We choose a particular static solution $\bar{q}$ and expand the field
around it, $q \rightarrow \bar{q} + q$. The linearised
equations for the fluctuations read
\begin{equation} 
\left( \bar{V}_{st} - \omega^2 \bar{g}_{st} \right) q^t = 0 ,
\end{equation}
where
\begin{equation}
\bar{g}_{st,u...v} = 
	\left. \partial_u \cdots \partial_v g_{st} \right|_{q = \bar{q}}
	\; , \;\;\;\;\;
	\bar{V}_{u...v} = \left.  
	\partial_u \cdots \partial_v V \right|_{q = \bar{q}} .
\end{equation}

The set of infinitesimal transformations $\{ \delta_a \}$ is split into
two sets, $\{ \delta_{a'} \}$ which change the static solutions, and
$\{ \delta_{\bar a} \}$ which leave them invariant.  The elements of
$G$ which leave the minimum invariant form a compact subgroup $H$ of
$G$ with Lie algebra ${\bf h}$ \cite{Fr85}. The parametrization
$\alpha$ can be chosen such that the generators of ${\bf g}-{\bf h}$
and ${\bf h}$, $\{ \delta_{a'} \}$ and $\{ \delta_{\bar a} \}$
respectively, are ortogonal to each other. Furthermore, we make
the simplifying assumption that $G/H$ is a symmetric space which is
valid for the Skyrme and $O(3)$ models. These conditions are resumed
in the following relations:
\begin{eqnarray}
{[}{\bf g}-{\bf h},{\bf h}]   &\subset & {\bf g}-{\bf h} ,\\
{[}{\bf h},{\bf h}]     &\subset & {\bf h}   ,\\
{[}{\bf g}-{\bf h},{\bf g}-{\bf h}] &\subset & {\bf h} ,
\end{eqnarray}
Therefore, the only non-zero structure constants are
${C_{a'\bar{b}}}^{c'}$, ${C_{\bar{a}\bar{b}}}^{\bar{c}}$ and
${C_{a'b'}}^{\bar{c}}$.

The generators of ${\bf g}-{\bf h}$ give rise to zero modes
\begin{eqnarray}
\bar{V}_{st} \psi^t_{a'} &=& 0 , \\
\psi^s_{a'} &=& \Im_{a'}^{-1/2} \delta_{a'} \bar{q}^s, 
	\label{zmodes} \\
\Im_{a'} &=& \bar{g}_{st} \delta_{a'} \bar{q}^s \delta_{a'} \bar{q}^t .
	\label{inert}
\end{eqnarray}

The set of normal modes $\{ \psi^s_n \} = \{ \psi^s_{\bar{n}} ;
\psi^s_{a'} \}$,
\begin{equation} 
\bar{V}_{st} \psi^t_{a'} = 0  \; , \;\;\;\;\;
	\left( \bar{V}_{st} - \omega_{\bar{n}}^2 \bar{g}_{st} \right) 
	\psi^t_{\bar{n}} = 0 
	\;\;\; (\omega_{\bar{n}} \neq 0), \label{nmodes}
\end{equation}
satisfies the orthogonality and completeness relations:
\begin{eqnarray}
& & \bar{g}_{st} \psi_{\bar{n}}^s \psi_{\bar{m}}^t = 
	\delta_{\bar{n}\bar{m}}
	\; , \;\;\;\;\;
	\bar{g}_{st} \psi_{a'}^s \psi_{b'}^t = \delta_{a'b'}
	\; , \;\;\;\;\;
	\bar{g}_{st} \psi_{\bar{n}}^s \psi_{a'}^t = 0 , \\
& & \sum_{\bar{n}} \bar{g}_{st} \psi_{\bar{n}}^t \psi_{\bar{n}}^u +
	\sum_{a'} \bar{g}_{st} \psi_{a'}^t \psi_{a'}^u = 
	\delta_s^u . \label{orthog}
\end{eqnarray}

We have assumed that the group  $G$ is unbroken at quantum level, so
the vacuum must be invariant under $G$. Since the static solution is
not, we restore the symmetry by the usual procedure of including
collective coordinates.

Now we perform a time-dependent transformation of the fields
$q$\cite{HK75,BK90}
\begin{equation} 
q^s \rightarrow R^s(q,\alpha(t)) . \label{trans}
\end{equation}
Care must be taken when performing this kind of transformations in the
path integral because extra terms of order $\hbar^2$ may
arise\cite{EG64}. With the choice of gauge we make below they are
absent, but in other gauges must be included\cite{CL80}. In the context
of collective coordinates quantization these terms were discussed
in\cite{GJ76}.

The transformed action reads
\begin{equation}
S'_0 = \int d\tau \left[ \frac{1}{2} g_{st} 
	\left( \dot{q}^s + 
	\sum_{a,b} \dot{\alpha}_a \zeta_{ab} \delta_b q^s \right)  
	\left( \dot{q}^t + 
	\sum_{c,d} \dot{\alpha}_c \zeta_{cd} \delta_d q^t \right) +
	V[q] \label{L'} \right] ,\label{S'}
\end{equation}
where the matrix $\zeta_{ab}$ is defined by means of the identity
\begin{equation}
\frac{\partial q^s}{\partial q'^t} \partial_a R^t(q,\alpha) = 
	\sum_b \zeta_{ab}(\alpha) \partial_b R^s(q,0) ,
\end{equation}
where $q'^s = R^s(q,\alpha)$. This is to be interpreted as follows: the
inverse of  $R$ maps $q'$ into $q$, and the differential of $R^{-1}$,
maps the tangent space to $M$ in $q'$ ($TM_{q'}$) into the tangent
space in $q$ ($TM_q$).  $ \partial_a R^s(q,\alpha) $ belongs to
$TM_{q'}$ and is mapped into $\zeta_{ab}\partial_b R^s(q,0)$ which
belongs to $TM_q$.

We now consider the parameters $\alpha_a$ as genuine (collective)
variables of the problem. It is easy to show that the action $S_0'$ is
invariant under gauge transformations \cite{HK75,BK90}
\begin{equation}
\left( \begin{array}{c} q^s \\ \alpha_b \end{array} \right)
	\rightarrow
	\exp{\sum_a \epsilon_a(\tau) \delta_a} 
	\left( \begin{array}{c} q^s \\ \alpha_b \end{array} \right) ,
	\label{gtrans}
\end{equation}
where 
\begin{equation}
\delta_a \alpha_b = - (\zeta^{-1})_{ab}.
\end{equation}

The above system may be interpreted as describing the problem from a
moving frame of reference oriented according to the collective
variables, and moving with velocity $\dot{\alpha}_a \zeta_{ab}$ along
the direction $b$. The gauge invariance is a manifestation of the fact
that transforming the intrinsic coordinates and correspondingly moving
the frame of reference leaves the physical situation unchanged.

\section{Antifield formalism} \label{AF}

The antifield formalism or Batalin-Vilkovisky formalism \cite{BV81}
provides a general method for the quantization of gauge theories within
a Lagrangian framework. The BRST transformation \cite{BRST74} and a
canonical formalism are defined in the space of Lagrangian variables by
the introduction of ghosts and antifields. A gauge-fixed quantum action
is obtained from an equation (Master Equation) plus boundary
conditions. The procedure is algebraic and straightforward, but
nevertheless powerful. The antifield formalism is reviewed in
\cite{HT92,GPS94}. 

In this section we apply the antifield formalism to the problem
presented in the previous sections. We start from the action
$S'_0[q,\alpha]$ (\ref{S'}) which is invariant under gauge
transformations (\ref{gtrans}). The set of variables $\{ q^s, \alpha_a
\}$ is enlarged by the introduction, for each of the independent gauge
generators, of (antihermitian) bosonic variables $b_a$, and (hermitian)
fermionic ghosts pairs $\eta_a$ and $\bar{\eta}_a$. The set of all
fields and coordinates will be generically denoted as $\{\phi_A\} =
\{q^s, \alpha_a, b_a,\eta_a,  \bar{\eta}_a \}$.

For each of the fields\footnote{Following the literature we denote such
quantities by fields, although some of them are actually variables in 
our case.}  
$\{\phi_A\}$ an {\em antifield} $\{\phi_A^*\}$ is introduced,
with opposite Grassmann parity and hermiticity.  The doubling of the
fields allows for the definition of a bracket structure, called the
antibracket. The antibracket of arbitrary functionals of the fields and
antifields is defined as
\begin{equation}
\left( F , G \right) = \int d\tau \left(
        \frac{\delta^R F}{\delta \phi_A(\tau)}
        \frac{\delta^L G}{\delta \phi_A^*(\tau)} -
        \frac{\delta^R F}{\delta \phi_A^*(\tau)}
        \frac{\delta^L G}{\delta \phi_A(\tau)} \right),
\end{equation}
where the superscript $R$ ($L$) stands for right (left) derivative which 
differ when deriving respect to a fermion. The
antibracket of fields and antifields plays an analogous role to the
Poisson bracket, because within this structure fields are conjugate to
antifields.

The next step is to find an action $S[\phi_A,\phi^*_A]$ which satisfies
the Master Equation $(S,S)=0$ with the boundary condition that it becomes
equal to the original action when the antifields vanish, i.e.,
$S[\phi_A,\phi^*_A=0]=S_0'[q,\alpha]$.  It can be
shown \cite{HT92,GPS94} that this requirements are satisfied by\footnote{To
check $(S,S)=0$ use
is  made of the group identity $
\frac{\partial}{\partial \alpha_c} \zeta_{ab} -
        \frac{\partial}{\partial \alpha_a} \zeta_{cb} +
         \zeta_{af} \zeta_{cd} C_{fd}{}^b = 0$.}
\begin{equation}
S = S' + \int d\tau \left(
         \sum_a q_s^* \delta_a q^s \eta_a -
         \sum_{ab}  \alpha^*_b \left(\zeta^{-1}\right)_{ab} \eta_a +
         \sum_{abc} \frac{1}{2} C_{ab}{}^c \eta^*_c \eta_a\eta_b +
         \sum_a \bar{\eta}^*_a b_a
         \right) .
\end{equation}
$S$ carries the information relative to the gauge algebra of the
problem through the presence of the structure constants. It is the
starting action to quantize the theory, for which a gauge fixing
procedure must be implemented.

The gauge fixed action is defined as
\begin{equation}
S_\psi \equiv 
	S\left[ \phi_A, \phi_A^*=\frac{\delta\psi}{\delta\phi_A}\right] ,
\end{equation}
where $\psi$ is an imaginary fermionic function of the fields
only.\footnote{$\psi$ is called the gauge fixing fermion.}
The path integral for the quantization of the classical theory
$S'$ is given by
\begin{equation}
Z = \int {\cal D}[q,\eta,{\bar \eta},b] 
	{\cal D}[\alpha] |\zeta| \sqrt{g}
        \exp(- S_\psi[q^s, \alpha_a, \eta_a, \bar{\eta}_a, b_a]) ,
\end{equation}
where ${\cal D}[\alpha] |\zeta|$ is the measure over the group
manifold, and ${\cal D}[q]\sqrt{g} = {\cal D}[q]\sqrt{\det(g_{st})}$ is
the measure over $M$, which can be exponentiated as
\begin{equation}
\int {\cal D}[q]\sqrt{g} e^{-S_{\psi}} = 
	\int {\cal D}[q]\prod_{\tau}\sqrt{g(\tau)} e^{-S_{\psi}} = 
	\int {\cal D}[q] 
	\exp\left(\frac{\delta(0)}{2}\int d\tau g(\tau)-S_{\psi}\right) 
	\label{Sg} .
\end{equation}

We make the following choice for the gauge fixing fermion
\begin{equation}
\psi = - i \int 
        \left\{
	  \sum_{a'} 
	  \left[
	    \left( 
              \omega^2_{a'} G_{a'} +
              \sum_b \frac{\partial}{\partial\tau}
              \left( \dot{\alpha}_b \zeta_{ba'} \right)
	    \right) + 
	    \frac{i \omega^2_{a'}}{2\Im_{a'}} b_{a'} 
	  \right] 
          \bar{\eta}_{a'} +
	  \sum_{\bar{a},b}
	  \dot{\alpha}_b \zeta_{b\bar{a}} 
	  \bar{\eta}_{\bar a}
	\right\} , \label{rho}
\end{equation}
where the functions $G_{a'}$ depend on the fields $q$, and the inertia
parameters $\Im_{a'}$ have been defined in (\ref{inert}). The
parameters $\omega_{a'}$ are arbitrary and should disappear from any
physical (gauge-invariant) result. They will be interpreted as the
frequencies associated with the spurious sector
(cf.\ Eqs.(\ref{Gsp}-\ref{Ggh})). The gauge fixed action reads
\begin{eqnarray}
S_{\psi} &=& S' + \int \left\{ 
	\sum_{a'} \left[
	\frac{\omega_{a'}^2}{2 \Im_{a'}} b_{a'} b_{a'} - 
	i b_{a'} \left( \omega^2_{a'} G_{a'} +
	\sum_b \frac{\partial}{\partial\tau} 
	\left( \dot{\alpha}_b \zeta_{ba'} \right) 
	\right) -
	i \dot{\bar{\eta}}_{a'} \dot{\eta}_{a'} 
	\right] 
	\right. \nonumber \\
	& & + 
	\sum_{\bar a} \left[
	- i b_{\bar a} 
	\left( \sum_b \dot{\alpha}_b \zeta_{b\bar{a}} \right) +
	i \bar{\eta}_{\bar a} \dot{\eta}_{\bar a} 
	\right] -
	\sum_{a',b}  
	i \omega_{a'}^2 \bar{\eta}_{a'} 
	\left( \delta_b G_{a'} \right) \eta_b 
	\nonumber \\
	& & \left. - 
	\sum_{a',b,c'} i C_{a' b}{}^{c'} 
	\left( \sum_d \dot{\alpha}_d \zeta_{da'} \right) 
	\eta_b \dot{\bar{\eta}}_{c'} +
	\sum_{a,b,\bar{c}} i {C_{a b}}^{\bar c} 
	\left( \sum_d \dot{\alpha}_d \zeta_{da} \right) 
	\eta_b \bar{\eta}_{\bar c} 
	\right\} .
\end{eqnarray}

The $b$ fields can be integrated out. Integration over $b_{\bar{a}}$
gives delta functions, and those over $b_{a'}$ are gaussian which can
be interpreted as averages over gauge fixing delta functions:
\begin{eqnarray}
Z &=& \int {\cal D}[q,\alpha,\eta,\bar{\eta}]  |\zeta| \sqrt{g}
	\prod_{\bar a} 
	\delta \left( \sum_b \dot{\alpha}_b \zeta_{b\bar{a}} \right) 
	\exp{( -S_\psi )} , \\
S_{\psi} &=& S'+ \int \left\{ 
	\sum_{a'} \left[
	\frac{\Im_{a'}}{2 \omega_{a'}^2}  
	\left( \omega_{a'}^2 G_{a'} + 
	\sum_b \frac{\partial}{\partial\tau} 
	\left( \dot{\alpha}_b \zeta_{ba'} \right) \right)
	\times \right. \right. \nonumber \\ 
        && \left. \left. \times 
	\left( \omega_{a'}^2 G_{a'} + 
	\sum_b \frac{\partial}{\partial\tau} 
	\left( \dot{\alpha}_b \zeta_{ba'} \right) \right) -
	i \dot{\bar{\eta}}_{a'} \dot{\eta}_{a'} 
	\right] + 	
	\sum_{\bar a} i \bar{\eta}_{\bar a} \dot{\eta}_{\bar a} 
	\right. \nonumber \\
	& & \left. -  
	\sum_{a',b}
	i \omega_{a'}^2 \bar{\eta}_{a'} 
	\left( \delta_b G_{a'} \right) \eta_b - 
	\sum_{a',b,c',d} i {C_{a' b}}^{c'} 
	\left( \dot{\alpha}_d \zeta_{da'} \right) 
	\eta_b \dot{\bar{\eta}}_{c'} +
	\sum_{a,b,\bar{c},d} i {C_{a b}}^{\bar c} 	
	\left( \dot{\alpha}_d \zeta_{da} \right) 
	\eta_b \bar{\eta}_{\bar c} 
	\right\} .
\end{eqnarray}

To get rid of the second order derivatives $\ddot{\alpha}$ let us
introduce auxiliary coordinates $\lambda_{a'}$ in the path integral for
$S_\psi$
\begin{eqnarray}
Z &=& \int {\cal D}[q,\alpha,\eta,\bar{\eta}]  |\zeta| \sqrt{g}
	\prod_{\bar a} 
	\delta \left( \sum_b \dot{\alpha}_b \zeta_{b\bar{a}} \right) 
	\exp{( -S_\psi )}
	\nonumber \\
	&=& 
	\int {\cal D}[q,\alpha,\eta,\bar{\eta},\lambda]  |\zeta| \sqrt{g}
	\prod_{a'} \delta \left(
	\frac{\omega_{a'}}{\sqrt{\Im_{a'}}}\lambda_{a'}
      - \sum_b \dot{\alpha}_b \zeta_{ba'} \right) 
	\prod_{\bar a} 
	\delta \left( \sum_b \dot{\alpha}_b \zeta_{b\bar{a}} \right) 
	\exp{( -S_\psi )}
\end{eqnarray}

The delta functions can be exponentiated by means of fields
${\cal P}$. We obtain
\begin{eqnarray}
Z &=& \int {\cal D}[q,\alpha,{\cal P},\lambda,\eta,\bar{\eta}] \sqrt{g} 
	\exp{( -S_\psi )} , \label{Zlag} \\
S_\psi' &=& S_{\rm intr.} + S_{\rm coll.} + S_{\rm coup.}  \label{Slag} ,\\
S_{\rm intr.} &=& \int d\tau \left[ \frac{1}{2} g_{st} 
     	\left( \dot{q}^s + \sum_{a'} 
     	\frac{\omega_{a'}}{\sqrt{\Im_{a'}}}\lambda_{a'}
     	\delta_{a'} q^s \right)  
     	\left( \dot{q}^t + \sum_{b'} 
     	\frac{\omega_{b'}}{\sqrt{\Im_{b'}}} \lambda_{b'}
     	\delta_{b'} q^t \right) +
    	 V[q] \right. \nonumber \\
 	&&- \sum_{a'} \frac{\Im_{a'}}{2 \omega_{a'}^2}  
     	\left( \omega_{a'}^2 G_{a'} -
	\frac{\omega_{a'}}{\sqrt{\Im_{a'}}}\dot{\lambda}_{a'}\right) 
   	\left( \omega_{a'}^2 G_{a'} -
	\frac{\omega_{a'}}{\sqrt{\Im_{a'}}}\dot{\lambda}_{a'}\right) 
	\nonumber \\ 
 	&&- \sum_{a'} i \dot{\bar{\eta}}_{a'} \dot{\eta}_{a'} + 	
     	\sum_{\bar a} i \bar{\eta}_{\bar a} \dot{\eta}_{\bar a} - 
    	\sum_{a',b} 
   	i \omega_{a'}^2 \bar{\eta}_{a'} 
        \left( \delta_b G_{a'} \right) \eta_b      	  
        \nonumber \\ && \left. - 
     	\sum_{a',b,c'} 
    	i {C_{a' b}}^{c'} \frac{\omega_{a'}}{\sqrt{\Im_{a'}}}\lambda_{a'} 
	\eta_b 	\dot{\bar{\eta}}_{c'} +
     	\sum_{a',b,\bar{c}}
    	i {C_{a' b}}^{\bar c} \frac{\omega_{a'}}{\sqrt{\Im_{a'}}}
        \lambda_{a'} 
	\eta_b \bar{\eta}_{\bar c} \right] \\
S_{\rm coll.} &=& - i \int d\tau \sum_a \dot{\alpha}_a {\cal P}_a
        \label{Scoll}  , \\
S_{\rm coup.} &=& \frac{i \omega}{a} \int d\tau \sum_{a'b}
        \frac{\omega_{a'}}{\sqrt{\Im_{a'}}}\lambda_{a'} 
	(\zeta^{-1})_{a'b} {\cal P}_b . \label{Scoup}
\end{eqnarray}

The action for the {\em intrinsic} variables (original fields $q$,
Lagrange multipliers $\lambda$ and ghosts $\eta,\bar{\eta}$) is $S_{\rm
intr.}$. It contains the transformed action (\ref{S'}), but with the
Lagrange multipliers as the velocities of the moving frame. It also has
a gauge fixing term, the action for the ghosts, and the coupling between
the ghosts and the bosonic variables.

$S_{\rm coll.}$ is the free action for the collective coordinates, in
Hamiltonian form. From it we see that ${\cal P}_a$ = $\delta
S_{coll.}/\delta {\dot \alpha_a}$ is to be interpreted as the canonical
momenta conjugate to the angles $\alpha_a$. Therefore ${\cal
D}[\alpha,{\cal P}]$ is the canonical invariant phase space measure.
The mixed Lagrangian-Hamiltonian form is frequent in collective
coordinate problems \cite{MZ94}.

The coupling between collective and intrinsic degrees of freedom is
given by $S_{\rm coup.}$.

\section{The Hamiltonian BRST Quantization} \label{HBQ}

In the previous section the gauge-fixed BRST Lagrangian path integral
was constructed by means of the antifield formalism. A similar result
can be obtained by integrating out the momenta in a Hamiltonian BRST
path integral. In this section we outline such a procedure to show its
equivalence to the antifield approach. Hamiltonian BRST quantization is
reviewed in Ref.\cite{HT92}. Its application to collective coordinates
can be found in \cite{BK90}.

The classical Hamiltonian corresponding to the gauge invariant
Lagrangian action (\ref{S'}) is \begin{equation} H = \frac{1}{2} g^{st}
p_s p_t + V[q] \label{H} , \end{equation} plus a set of abelian
first-class constraints $F_a = \sum_b \zeta_{ab} j_b - {\cal P}_a$,
with Poisson brackets $\{ p_s , q^t \} = - \delta_s^t$, $\{ {\cal P}_a
, \alpha_b \} = - \delta_{ab}$, and $j_a = p_s \delta_a q^s$. The
quantum Hamiltonian has an ordering ambiguity which is saved as usual
by taking the kinetic part to be the laplacian over $M$.  A new set of
non-abelian constraints can be obtained by multiplying $F$ by
$\zeta^{-1}$ \cite{BK90}
\begin{equation}
f_a = \sum_b (\zeta^{-1})_{ab} F_b = j_a - I_a 
	\; , \;\;\;\;\;
	\{ f_a , f_b \} = - \sum_c {C_{ab}}^c f_c . \label{algf}
\end{equation}
The operators $I_a = \sum_b (\zeta^{-1})_{ab} {\cal P}_b$ satisfy $\{
I_a , I_b \} = \sum_c {C_{ab}}^c I_c$, and should be considered as the
collective version of the intrinsic operators $j_a$. 

The phase space is enlarged by introducing, for each first-class
constraint, a Lagrange multiplier $\lambda_a$ and two (fermionic)
ghosts $\eta_a, {\bar \eta}_a$, together with their corresponding
conjugate momenta $b_a,\pi_a,{\bar \pi}_a$.  Upon quantization,
physical states are defined as those annihilated by the BRST charge
$\Omega$
\begin{equation}
\Omega \equiv \sum_a \left( b_a \bar{\pi}_a - 
	f_a \eta_a \right) -
	\frac {i}{2} \sum_{a,b,c} {C_{ab}}^c \eta_a \eta_b \pi_c , 
\end{equation}
and physical operators as those that commute with $\Omega$.

Any term of the form $[ \rho,\Omega ]$ may be
added to the Hamiltonian without altering the overlaps of the original
Hamiltonian within the subspace annihilated by $\Omega$. After rescaling the variables: $\lambda_{a'}
\rightarrow i \omega_{a'} \Im_{a'}^{-1/2} \lambda_{a'}$, $b_{a'}
\rightarrow - i \Im_{a'}^{1/2} \omega_{a'}^{-1} b_{a'}$, we choose
\cite{BK90}
\begin{equation}
\rho= \sum_{a'} \left[ \frac{i \omega_{a'}}{\sqrt{\Im_{a'}}} 
	\lambda_{a'} \pi_{a'} + 
	\left( \omega^2_{a'} G_{a'} + \frac{i}{2} b_{a'} \right)
	\bar{\eta}_{a'} \right],	
\end{equation} 
which yields, after integrating the momenta $p$, $b$, $\pi$ and
$\bar{\pi}$ in the phase space path integral, the Lagrangian path
integral of Eqs.(\ref{Zlag})-(\ref{Slag}).

The functions $G$ are chosen such that $[ j_{a'} , G_{b'} ] \neq
0$ at leading order, so that the ghost propagators are well-defined. In
that case, the BRST Hamiltonian does not commute with the
operators $j_a$. This is desirable for the $j_{a'}$'s, since they give
rise to the zero modes, as seen in Sec.\ref{SSCC}. On the other hand, the
symmetry under the transformations $j_{\bar a}$ was not broken by the
classical solution, and it is convenient to choose a gauge-fixing
scheme that conserves such a symmetry. 
First we must 
 define transformation operators similar to $f_a$ (which
transform the original and collective variables) for the Lagrange
multipliers and the ghosts \cite{BK90}:
\begin{eqnarray}
N_a &\equiv& - \sum_{bc} {C_{ab}}^c 
	\frac{\omega_{b'} \sqrt{\Im_{c'}}}{\omega_{c'} \sqrt{\Im_{b'}}}
	\lambda_b b_c , \\
\tau_a &\equiv& i \sum_{bc} {C_{ab}}^c \eta_b \pi_c , \\
\bar{\tau}_a &\equiv& i \sum_{bc} {C_{ab}}^c \bar{\pi}_b \bar{\eta}_c .
\end{eqnarray}
The above operators have the same commutation relations as the $j$'s.

Transformations of all the variables are generated by the operators
\begin{eqnarray}
L_a &\equiv& f_a + N_a + \tau_a + \bar{\tau}_a , \\
{[} L_a , L_b ] &=& - \sum_{bc} {C_{ab}}^c L_c ,
\end{eqnarray}
which are null operators \cite{BK90}
\begin{equation}
L_a = [ \Omega , - \sum_{bc} {C_{ab}}^c 
	\frac{i \omega_{b'}}{\sqrt{\Im_{b'}}} \lambda_b \bar{\eta}_c 
	- \pi_a ] , 
\end{equation}
so they commute with $\Omega$, and map physical states into null (zero
norm) states \cite{HT92}. They also commute with the original
Hamiltonian.

The operators $L_{a'}$ do not commute with the gauge-fixing fermion due
to the functions $G_{a'}$. However, this functions can be chosen to
transform under the $L_{\bar a}$'s as,
\begin{equation}
i [ L_{\bar a} , G_{b'} ] = i [ j_{\bar a} , G_{b'} ] = 
	- {C_{\bar{a}c'}}^{b'} G_{c'} .
\end{equation}
This is indeed the choice made in Sec.\ref{PT} (Eq.\ref{G}). 

If the arbitrary parameters $\omega_{a'}$ are all taken
equal\footnote{In fact they need only be taken equal within each
irreducible representation of $H$ in ${\bf g} - {\bf h}$} then the
operators $L_{\bar a}$ commute with the BRST Hamiltonian.  Thus the
eigenstates of the Hamiltonian can be classified by irreducible
representations of the group $H$.  However, in an irreducible
representation any state can be obtained from any other by repeated
application of the $L_{\bar{a}}$'s, which are null operators.  This
means that the only representation which is not neccesarily composed by
null-states is the trivial one in which there is only one state
satisfying $L_{\bar{a}}|\rangle = 0$. Physical states belong to this
representation.

The action of collective operators $I_{\bar a}$ on physical states is
determined by the intrinsic structure, i.e.
\begin{equation}
L_{\bar a} | \mbox{ph} \rangle = 0 \Rightarrow
	I_{\bar a} | \mbox{ph} \rangle =
	( j_{\bar a} + N_{\bar a} + \tau_{\bar a} + \bar{\tau}_{\bar a} ) 	
	| \mbox{ph} \rangle .
\end{equation}
Thus the collective operators $I_{\bar a}$ can be identified with the
intrinsic operators $j_{\bar a} + N_{\bar a} + \tau_{\bar a} +
\bar{\tau}_{\bar a}$ when taking matrix elements between physical
states. This fact is used in the next section when the partition
function of the system is calculated.

\section{The Perturbative Treatment} \label{PT}

In this section we evaluate perturbatively the partition function
(\ref{Zlag}), i.e., the trace over physical states of $e^{-\beta H}$.
It is known \cite{HK80} that this amounts to perform the functional
integration over fields with periodic boundary conditions in the
interval $[0,\beta]$, including the ghosts despite the fact that they
are fermions.

Minimization of the action (\ref{Slag}) with the assumption that
Lagrange multipliers and ghosts vanish classically gives that the
only non-vanishing expectation values are those of Eq.(\ref{min}).
Expanding all the fields as fluctuations around their classical values
allows to perform a perturbative calculation. The quadratic Lagrangian
provides the free propagators, and the third or higher order terms give
vertices to be used in the usual Feynman diagram expansion.

To ease the calculation we expand the fluctuations of $q^s$ in terms of the
normal modes
\begin{equation}
q^s = \bar{q}^s + \sum_{\bar{n}}  \xi_{\bar{n}}  \psi_{\bar{n}}^s
                + \sum_{a'}       \xi_{a'}       \psi_{a'}^s ,
\end{equation}
where $\xi_n$ are normal coordinates.
We choose the gauge fixing functions $G_{a'}$ to be
\begin{equation}
G_{a'} = \Im_{a'}^{-1/2} g_{st} \psi^s_{a'} (q^t - \bar{q}^t) 
       = \Im_{a'}^{-1/2} \xi_{a'} . \label{G}
\end{equation}
Such choice cancels at quadratic level the coupling between $\lambda$
and $\xi$, and provides (spurious) frequencies to the zero modes and to
the ghosts. This gauge is known as 't Hooft gauge \cite{IZ80}, or as
rigid gauge in soliton problems.

Now the quadratic action can be written,
\begin{eqnarray}
S^{(2)}_{\rm int.} &=& \int d\tau \left( 
	\sum_{\bar{n}} \frac{1}{2} \left( 
	\dot{\xi}_{\bar{n}} \dot{\xi}_{\bar{n}} + 
	\omega_{\bar{n}}^2 \xi_{\bar{n}} \xi_{\bar{n}}
	\right)    
        + \sum_{a'} \frac{1}{2} \left( 
	\dot{\xi}_{a'} \dot{\xi}_{a'} + \omega_{a'}^2 \xi_{a'} \xi_{a'} 
  	+ \dot{\lambda}_{a'} \dot{\lambda}_{a'} + 
	\omega_{a'}^2 \lambda_{a'} \lambda_{a'} 
	\right) \right.
	\nonumber \\
        && \left.
        - i \sum_{a'} \left( 
	\dot{\bar{\eta}}_{a'} \dot{\eta}_{a'} + 
	\omega_{a'}^2 \bar{\eta}_{a'} \eta_{a'} 
	\right) 
        + i \sum_{\bar{a}}
	\bar{\eta}_{\bar{a}} \dot{\eta}_{\bar{a}}
        \right) . \label{S2}
\end{eqnarray}

To evaluate the partition function to leading order we use the gaussian
integrals over periodic time $\tau$
\begin{equation}
\begin{array}{lclc}
\int {\cal D}[\phi] 
	\exp{\left[-\frac{1}{2}\int d\tau (\dot{\phi}^2  +
	\omega^2 \phi^2)\right]}
  	&=& {\det}^{-1/2}(\partial_{\tau}^2-\omega^2) &
  	\mbox{(bosonic)} ,\\
\int {\cal D}[\bar{\eta}\eta] 
	\exp{\left[-\int d\tau (\dot{\bar{\eta}}\dot{\eta}  +
     	\omega^2 \bar{\eta}\eta)\right]} 
	&=& {\det}(\partial_{\tau}^2-\omega^2) &
  	\mbox{(fermionic)} .
\end{array}
\end{equation}
We obtain \cite{Ka93}:
\begin{eqnarray}
Z^{(1)} &=& \int
        {\cal D}[\xi_n,\lambda_{a'},\bar{\eta}_{a'},\eta_{a'}] \sqrt{g}
        \exp(-S_{\rm intr.}^{(2)}) \nonumber \\
    &=& \prod_{\bar{n}} {\det}^{-1/2}(\partial^2_{\tau}-\omega^2_{\bar{n}})
        \prod_{a'}{\det}^{-1/2}(\partial^2_{\tau}-\omega^2_{a'})
                  {\det}^{-1/2}(\partial^2_{\tau}-\omega^2_{a'})
                  \det(\partial^2_{\tau}-\omega^2_{a'}) \nonumber \\
    &=& \prod_{\bar{n}} \left[ 2 \sinh 
	\left( \frac{\beta \omega_{\bar{n}}}{2} \right) \right]^{-1} ,
\end{eqnarray}
where $\beta$ is the inverse temperature. As usual, the ghost
determinant cancels out the contribution from the bosonic spurious
degrees of freedom $(\lambda_{a'},\xi_{a'})$.  Expanding the above
result the usual contribution $\sum_{\bar{n}} \frac{\omega_{n}}{2}$ to
the vacuum energy is found.

The finite temperature propagators can be similarly calculated
\begin{eqnarray}
\langle \langle T \xi_{\bar{n}}(\tau) \xi_{\bar{n}}(\tau') \rangle \rangle
        &=& G_{\omega_{\bar{n}}}(\tau - \tau') ,\\
\langle \langle T \xi_{a'}(\tau) \xi_{a'}(\tau') \rangle \rangle
        &=& G_{\omega_{a'}}(\tau - \tau') , \label{Gsp} \\
\langle \langle T \lambda_{a'}(\tau)\lambda_{a'}(\tau') \rangle \rangle
        &=& G_{\omega_{a'}}(\tau - \tau') , \label{Glm} \\
\langle \langle T \bar{\eta}_{a'}(\tau)\eta_{a'}(\tau') \rangle \rangle
        &=& -i G_{\omega_{a'}}(\tau - \tau') \label{Ggh} , 
\end{eqnarray}
where
\begin{equation}
G_{\omega}(\tau) = \frac{1}{2 \omega \left( e^{\beta \omega} - 1 \right)}
        \left( e^{\beta \omega} e^{- \omega |\tau|} +
        e^{\omega |\tau|} \right) .
\end{equation}
The time order in the propagator for two velocities is defined as
\begin{eqnarray}
\langle \langle \hat{T} \dot{\xi}_n(\tau)
	\dot{\xi}_n(\tau') \rangle \rangle
	&=& \partial_{\tau}\partial_{\tau'}G_{\omega_n}(\tau - \tau')
	\nonumber \\
    	&=& \delta(\tau-\tau')-\omega^2_n G_{\omega_n}(\tau - \tau').
\end{eqnarray}
The $\delta(\tau-\tau')$ will cancel extra vertices in the Lagrangian
(respect to the interaction Hamiltonian), as well as the $\delta(0)$
contributions arising from the measure $\frac{1}{2}\delta(0)\int d\tau
g(\tau)$ (Eq.(\ref{Sg})) \cite{Ho72,We71}.

\begin{figure}
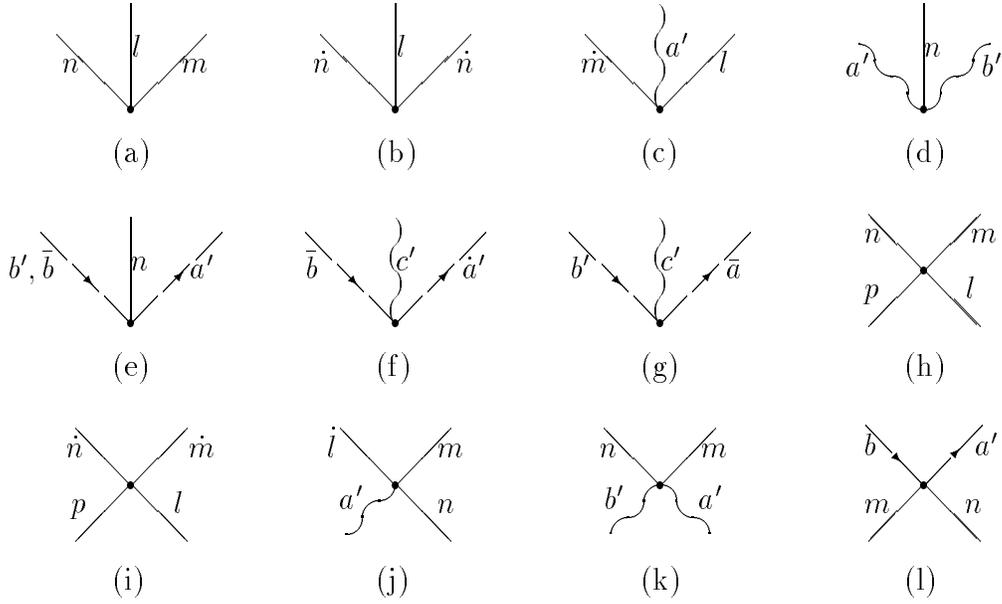

\caption{Third and fourth order vertices from $L_{\rm
BRST}^{(3)}$ and $L_{\rm BRST}^{(4)}$.} \label{fvert}
\end{figure}

The higher than quadratic terms in the Lagrangian provide the vertices
for constructing Feynman diagrams. The cubic and quartic vertices are
given in Fig.\ref{fvert}. The straight lines correspond to the fields
$q$, the wiggy lines to $\lambda$, and the broken lines to ghosts. The
values for the vertices of Fig.\ref{fvert} are (hereafter summation is
implicit over any repeated index):
\begin{eqnarray}
\mbox{(a)} &=& - \frac{1}{6} A_{nml} \xi_n \xi_m \xi_l , \\
\mbox{(b)} &=& - \frac{1}{2} B_{nm;l}\dot{\xi}_n \dot{\xi}_m \xi_l , \\
\mbox{(c)} &=& - \omega_{a'} \left( B_{a'm;n} + D^{a'}_{mn} \right) 
                 \lambda_{a'} \xi_n \dot{\xi}_m, \\
\mbox{(d)} &=& - \frac{1}{2} \omega_{a'}\omega_{b'} \left( B_{a'b';n} 
               + D^{a'}_{b'n} + D^{b'}_{a'n} \right) 
                  \lambda_{a'}\lambda_{b'}\xi_n , \\
\mbox{(e)} &=&  i \omega_{a'}^2 \sqrt{\frac{\Im_{b'}}{\Im_{a'}}}
               D^{b'}_{a'n} 
		\bar{\eta}_{a'} \xi_n \eta_{b'}
             + i \omega_{a'}^2 {\Im_{a'}}^{-\frac{1}{2}}
               D^{\bar{b}}_{a'n} 
		\bar{\eta}_{a'} \xi_n \eta_{\bar{b}}, \\
\mbox{(f)} &=& - i \omega_{c'}{\Im_{c'}}^{-\frac{1}{2}} 
		{C_{c'\bar{b}}}^{a'}
               \dot{\bar{\eta}}_{a'} \lambda_{c'} \eta_{\bar{b}}, \\
\mbox{(g)} &=&  i \omega_{c'}{\Im_{c'}}^{-\frac{1}{2}} \	
		{C_{c'b'}}^{\bar{a}}
               \bar{\eta}_{\bar{a}} \lambda_{c'} \eta_{b'} , \\
\mbox{(h)} &=& - \frac{1}{24} E_{nmlp} \xi_n \xi_m \xi_l \xi_p , \\
\mbox{(i)} &=& - \frac{1}{4}  F_{nm,lp} \dot{\xi}_n \dot{\xi}_m \xi_l
                 \xi_p ,\\
\mbox{(j)} &=& - \omega_{a'} \left(\frac{1}{2} F_{la';nm} + 
                 B_{pl;n} D^{a'}_{pm} + H^{a'}_{l,nm}\right) 
                 \lambda_{a'} \xi_n \xi_m \dot{\xi}_l , \\
\mbox{(k)} &=& - \omega_{a'}\omega_{b'} \left[ \left(
	\frac{1}{4} F_{a'b';nm} +
	B_{lb';n} + \frac{1}{2}
                 D^{a'}_{ln} \right) 
                 D^{b'}_{lm} + H^{a'}_{b';nm} \right]
                 \lambda_{a'}\lambda_{b'}\xi_n \xi_m , \\
\mbox{(l)} &=& \omega_{a'}^2 H^{b}_{a';nm} i \bar{\eta}_{a'} \eta_{b} 
               \xi_n \xi_m.
\end{eqnarray}
 
\begin{figure}
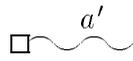

\caption{Collective-intrinsic coupling vertex.} \label{fvertj}
\end{figure}

There is also a coupling term between $\lambda$ and the collective
operators $I$ (cf.\ Eq.(\ref{Slag})). When integrating over the fields
$q$, $\lambda$ and the ghosts, the $I$'s behave like (non-commuting)
sources for $\lambda$. The corresponding vertex is shown in
Fig.\ref{fvertj}, and is equal to $-i \lambda_{a'}
I_{a'}\omega_{a'}/\sqrt{\Im_{a'}}$.

\subsection{Two-loop Corrections}

\begin{figure}
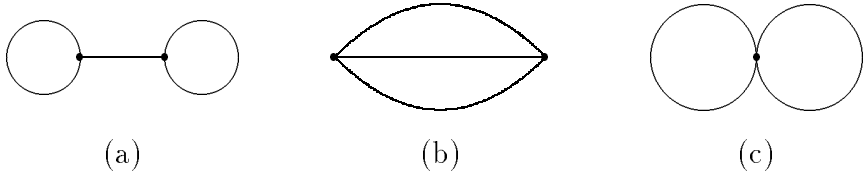

\caption{Two-loops diagrams involved in the correction to the partition
function.} \label{fdiags}
\end{figure}

The  two loops correction to the partition function (excluding the
terms depending on $I_{a'}$ which we evaluate below) is given by
the diagrams of Fig.\ref{fdiags}. The values of the diagrams are,
\begin{eqnarray}
\mbox{(a)} &=&  
	\frac{1}{8}  	
	\left[ \left( \omega_{\bar{m}}^2 B_{\bar{m}\bar{m};\bar{n}} 
      - A_{\bar{m}\bar{m}\bar{n}} \right) G_{\omega_{\bar{m}}}(0)   
      + \omega^2_{\bar{n}}   D^{a'}_{\bar{n}a'} 
        G_{\omega_{a'}}(0)\right]^2 
	g(\omega_{\bar n}) \\
\mbox{(b)} &=& 
        \frac{\beta}{2}\omega^2_{a'}\left(B_{a'm;n} + D^{a'}_{mn}\right)^2
         G_{\omega_{a'}}(0) G_{\omega_n}(0) \nonumber \\
&&      - \frac{\beta}{2}\omega^2_m B^2_{mp;n} 
         G_{\omega_m}(0) G_{\omega_n}(0) \nonumber \\
&&       - \frac{1}{2} \frac{\omega^2_{c'}}{\Im_{c'}}
         C^{a'}_{c'\bar{b}}C^{\bar{b}}_{c'a'} 
         g(\omega_{c'},\dot{\omega}_{a'}) \nonumber \\
&&    + \frac{1}{4} \omega^2_{a'}\omega^2_{b'}
         \left[ \left( B_{a'b';n} + D^{a'}_{b'n} + D^{b'}_{a'n}\right)^2
         - 2 D^{a'}_{b'n} D^{b'}_{a'n} \right] 
         g(\omega_{a'},\omega_{b'},\omega_n) \nonumber \\
&&    - \frac{1}{2} \omega^2_{a'}\omega^2_m 
         \left( B_{a'm;n} + D^{a'}_{mn} \right)^2
         g(\omega_{a'},\omega_m,\omega_n) \nonumber \\
&&    + \frac{1}{4} \left( \omega^2_m\omega^2_p B^2_{mp;n} +
         \frac{1}{3} A^2_{nmp}\right) 
         g(\omega_n,\omega_m,\omega_p) \nonumber \\
&&    - \frac{1}{2} \omega^2_{a'} \left(B_{a'm;n} + D^{a'}_{mn} \right)
         \left(B_{a'n;m} + D^{a'}_{nm} \right)
         g(\omega_{a'},\dot{\omega}_m,\dot{\omega}_n) \nonumber \\
&&    + \frac{1}{2} \left( \omega^2_n B_{mn;p} B_{pn;m} +
          B_{mp;n}A_{mpn} \right)
          g(\omega_n,\dot{\omega}_m,\dot{\omega}_p) \\
\mbox{(c)} &=& 
	\frac{\beta}{4} \omega^2_{\bar{n}} F_{\bar{n}\bar{n};ll} 
         G_{\omega_{\bar{n}}}(0)  G_{\omega_{l}}(0)                    
	\nonumber \\
   &&  - \beta \omega^2_{a'} 
         \left(B_{la';n} + \frac{1}{2} D^{a'}_{ln}\right) D^{a'}_{ln}
         G_{\omega_{a'}}(0)  G_{\omega_{n}}(0)                         
	\nonumber \\
   &&  - \frac{\beta}{8} E_{nnll}G_{\omega_{n}}(0)G_{\omega_{l}}(0)
\end{eqnarray}	
where the functions $g(\omega_{\bar{n}})$, 
$g(\omega_{\bar{n}},\omega_{\bar{l}},\omega_{\bar{m}})$,
$g(\omega_{\bar{l}},\dot{\omega}_{\bar{n}},\dot{\omega}_{\bar{m}})$
and $g(\omega_n,\dot{\omega}_m)$ are
defined in the Appendix. The contributions proportional to
$\delta(0)$ coming from the exponentiation of $\prod_{\tau}\sqrt{g}$
(cf.\ Eq.(\ref{Sg})) and from the propagator
$\langle\langle\hat{T}\dot{\xi}_n\dot{\xi}_n\rangle\rangle$ cancel
exactly (as expected) and are not listed.

Summing up all diagrams and using the relations given in the Appendix 
to simplify the result we obtain:
\begin{eqnarray}
Z^{(2)} &=& 	\frac{1}{4}  \left(\omega^2_{\bar{m}}\omega^2_{\bar{l}}
                     B^2_{\bar{m}\bar{l};\bar{n}} 
       + \frac{1}{3} A^2_{\bar{n}\bar{m}\bar{l}}\right) 
         g(\omega_{\bar{n}},\omega_{\bar{l}},\omega_{\bar{m}}) 
	\nonumber \\ &&
       + \frac{1}{2} 
         \left(B_{\bar{m}\bar{n};\bar{l}} A_{\bar{m}\bar{n}\bar{l}}
       + \omega_{\bar{l}} B_{\bar{m}\bar{l};\bar{n}} 
          B_{\bar{n}\bar{p};\bar{m}} \right) 
    g(\omega_{\bar{l}},\dot{\omega}_{\bar{n}},\dot{\omega}_{\bar{m}}) 
	\nonumber \\ 
&&     + \frac{1}{8}
	\left[ \left( \omega_{\bar{m}}^2 B_{\bar{m}\bar{m};\bar{n}} 
       - A_{\bar{m}\bar{m}\bar{n}} \right) G_{\omega_{\bar{m}}}(0)   
         \right]^2 g(\omega_{\bar n})
	\nonumber \\
&&     + \left( -\frac{1}{8} E_{\bar{n}\bar{n}\bar{m}\bar{m}} 
       + \frac{1}{4} \omega^2_{\bar{n}}F_{\bar{n}\bar{n};\bar{m}\bar{m}}
       - \frac{1}{2} \omega^2_{\bar{n}} B^2_{\bar{n}l;\bar{m}} \right)
          G_{\omega_{\bar{n}}}(0)  G_{\omega_{\bar{m}}}(0)  
	\nonumber \\
&&      - \frac{\beta}{8} 
         \left(2 B_{a'\bar{m}\bar{n}} + D^{a'}_{\bar{m}\bar{n}}\right)
         \left[D^{a'}_{\bar{n}\bar{m}} + 
	4 D^{a'}_{\bar{m}\bar{n}}\omega^2_{\bar{m}} \beta
          G_{\omega_{\bar{n}}}(0)  G_{\omega_{\bar{m}}}(0) \right] 
	\nonumber \\
&&     + \frac{\beta}{8} D^{b'}_{a'c'} D^{a'}_{b'c'} 
       + \frac{\beta}{8} D^{a'}_{b'\bar{m}} D^{b'}_{a'\bar{m}} 
       - \frac{\beta}{8} B_{a'b';\bar{m}} D^{a'}_{\bar{m}b'} 
	- \frac{\beta}{8\Im_{c'}} C^{a'}_{c'\bar{b}} C^{\bar{b}}_{c'a'} ,
\end{eqnarray}
where we see that all spurious frequencies have disappeared, as it
should. It can be checked that in simple cases the known results are
reobtained, as for example the $\lambda \phi^4$ kink in $1+1$
dimensions\cite{GJ76}.

\subsection{Collective Hamiltonian}

\begin{figure}
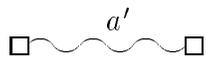

\caption{Collective energy to lowest order.} \label{fdiagj}
\end{figure}

The corrections involving the collective momenta $I_{a}$ can be
evaluated in a similar way as in the previous section.  However, as is
done in perturbation theory to a degenerate level the result is best
expressed as an effective Hamiltonian acting over collective variables.
To tree level the diagrams involved are the ones of Fig.\ref{fdiagj},
which are of ${\cal O}(\Im^{-1})$
\begin{equation}
H_{\rm coll} = \frac{1}{2 \Im_{a'}} I_{a'}^2 + {\cal O}(\Im^{-2}) .
\label{H2}
\end{equation}

\begin{figure}
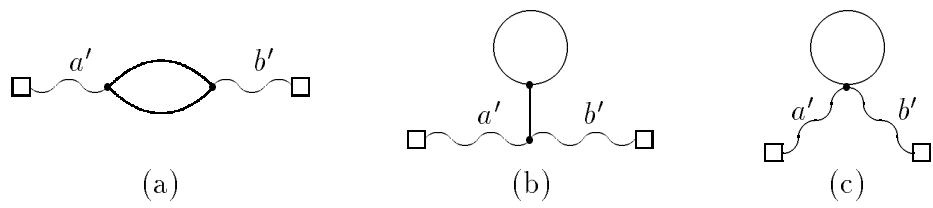

\caption{First correction to the collective energy.}
	\label{fdiagsj}
\end{figure}

To one-loop the diagrams are the ones of Fig.\ref{fdiagsj}. They give
\begin{eqnarray}
\mbox{(a)} &=& \frac{\{I_{a'},I_{b'}\}}{\sqrt{\Im_{a'}\Im_{b'}}}
      \left\{
      \frac{1}{4} \left(B_{a'm;n} + D^{a'}_{mn}\right)
      \left(B_{b'm;n} + D^{b'}_{mn}\right)
	 \right. \times \nonumber \\ && \times  
      \left[- \beta G_{\omega_{\bar{n}}}(0) 
    + \omega^2_m g(\omega_n,\omega_m) \right] 
	\nonumber \\ 
&&   + \frac{1}{4} \left(B_{a'm;n} + D^{a'}_{mn}\right)
      \left(B_{b'n;m} + D^{b'}_{nm}\right)
      g(\dot{\omega}_n,\dot{\omega}_m) 
	\nonumber \\
&&  \left. 
	- \frac{1}{4} \left(B_{a'c';n} + D^{a'}_{c'n}+D^{c'}_{a'n}\right)
     \left(B_{b'c';n} + D^{b'}_{c'n}+D^{c'}_{b'n}\right)
      \omega^2_{c'} g(\omega_{n},\omega_{c'}) \right\} 
	\nonumber \\
&& - \frac{\beta I_{a'} I_{b'}}{\Im_{a'}\Im_{b'}} 
     C^{d'}_{b'\bar{c}}C^{\bar{c}}_{a'd'} 
     G_{\omega_{d'}}(0) , \\
\mbox{(b)} &=& \frac{\{I_{a'},I_{b'}\}}{8 \sqrt{\Im_{a'}\Im_{b'}}}
      \left(B_{a'b';\bar{n}} + 2 D^{a'}_{b'\bar{n}} \right)
 	\times \nonumber \\ && \times
     \left[\left(\omega^2_{\bar{m}} B_{\bar{m}\bar{m};\bar{n}} 
    - A_{\bar{m}\bar{m}\bar{n}} \right) 
      G_{\omega_{\bar{m}}}(0)
    + \omega^2_{\bar{n}} D^{c'}_{\bar{n}c'} G_{\omega_{c'}}(0) 
	\right] g(\omega^2_{\bar{n}}), \\
\mbox{(c)} &=& \frac{\beta \{I_{a'},I_{b'}\}}{\sqrt{\Im_{a'}\Im_{b'}}}
       G_{\omega_{n}}(0) \left[ \frac{1}{8} F_{a'b';nn} 
     + \frac{1}{2} B_{la';n}D^{b'}_{ln} 
     + \frac{1}{4} D^{a'}_{ln} D^{b'}_{ln} 
     + \frac{1}{4}H^{a'}_{b';nn}\right] .
\end{eqnarray}
There is a further contribution from the vertex 
 $-i \lambda_{a'} I_{a'}\omega_{a'}/\sqrt{\Im_{a'}}$ (Fig.\ref{fvertj})
at fourth order. Such contribution correspond to disconnected diagrams and
should vanish but for the fact that the
generators $I_{a'}$ do not conmute. Using perturbation theory to a
degenerate level we obtain
\begin{equation}
-\frac{I_{a'}I_{a'}I_{b'}I_{b'}}{\Im_{a'}\Im_{b'}} \beta 
	\frac{1}{2}G_{\omega_{a'}}(0) + 
	\frac{I_{a'}I_{b'}I_{a'}I_{b'}}{\Im_{a'}\Im_{b'}} 
	\frac{1}{8} g(\dot{\omega}_{a'},\dot{\omega}_{b'}) -
	\frac{I_{a'}I_{b'}I_{b'}I_{a'}}{\Im_{a'}\Im_{b'}} 
	\left( \frac{1}{8} g(\dot{\omega}_{a'},\dot{\omega}_{b'}) - 
	\frac{1}{2}\beta G_{\omega_a}(0)\right) \label{4Is} ,
\end{equation} 
which clearly vanishes if the $I$'s commute. As was discussed in 
Sec.\ref{HBQ}, the $I_{\bar a}$'s can be identified with intrinsic operators,
so that the contributions proportional to them should be included in the
intrinsic correction to ${\cal O}(\Im^{-2})$ (three-loop diagrams). 
Taking this into account, the contribution of (\ref{4Is}) to the collective 
hamiltonian reads
\begin{equation}
\frac{I_{a'}I_{b'}}{\Im_{a'}\Im_{b'}} 
	\frac{1}{2} \beta G_{\omega_{c'}}(0)
	C_{c'b'}{}^{\bar{d}} C_{c'\bar{d}}{}^{a'}
	\label{4I} .
\end{equation}

The one-loop correction to the effective collective Hamiltonian (\ref{H2})
can be obtained as  $-1/\beta$ times
the sum of all diagrams plus (\ref{4I}). Using the identities of
the Appendix it reads:
\begin{eqnarray}
H^{1-{\rm loop}}_{\rm coll} &=& 
     - \frac{\{I_{a'},I_{b'}\}}{\sqrt{\Im_{a'}\Im_{b'}}}
       G_{\omega_{\bar{n}}}(0)
       \left[
        \frac{1}{8} F_{a'b',\bar{n}\bar{n}} +
        \frac{1}{4} H^{a'}_{b';\bar{n}\bar{n}}
	\right. \nonumber \\ &&  
     + \frac{1}{8\omega^2_{\bar{n}}}
       \left(B_{a'b';\bar{n}}+ 2 D^{a'}_{b'\bar{n}}\right)
       \left(\omega^2_{\bar{m}} B_{\bar{m}\bar{m};\bar{n}} -
	A_{\bar{m}\bar{m}\bar{n}}\right) \nonumber \\
&&   - \frac{1}{4} B_{a'm;\bar{n}}B_{b'm;\bar{n}} 
     - \frac{1}{4} D^{c'}_{a'\bar{n}}D^{c'}_{b'\bar{n}} 
	\nonumber \\ && \left. 
     - \frac{1}{2} \left(B_{a'c';\bar{n}} D^{c'}_{b'\bar{n}} 
     + D^{a'}_{c'\bar{n}}D^{c'}_{b'\bar{n}} \right) 
	\right] 
	\nonumber \\
&&   - \frac{\{I_{a'},I_{b'}\}}{\beta\sqrt{\Im_{a'}\Im_{b'}}}
       \left[
       \frac{1}{4} \omega^2_{\bar{m}}
       \left( B_{a'\bar{m};\bar{n}} + D^{a'}_{\bar{m}\bar{n}}\right)
       \left( B_{b'\bar{m};\bar{n}} + D^{b'}_{\bar{m}\bar{n}}\right)
       g(\omega_{\bar{n}},\omega_{\bar{m}}) \right. 
	\nonumber \\
&&   + \left. \frac{1}{4}  
       \left( B_{a'\bar{m};\bar{n}} + D^{a'}_{\bar{m}\bar{n}}\right)
       \left( B_{b'\bar{n};\bar{m}} + D^{b'}_{\bar{n}\bar{m}}\right)
       g(\dot{\omega}_{\bar{n}},\dot{\omega}_{\bar{m}}) \right] .
\end{eqnarray}
Again it is seen to be independent of the arbitrary parameters
$\omega_{a'}$, which gives a useful test of the calculation. It has 
a dependence in $\beta$ because it comprises the collective Hamiltonian
asocciated with all vibrational states. If, for example, we need the
collective energies associated with the ground state, we just extract
the $\beta \rightarrow \infty$ limit. If we need that of the 
$\omega_{\bar{n}}$ level, we look for the $e^{-\beta \omega_{\bar{n}}}$
dependent term.

The above result gives the correction to the inertia parameters $\Im_{a'}$. 
To obtain the correction to ${\cal O}(\Im^{-2})$ to the energy levels the 
three-loop intrinsic diagrams should be computed.

\section{Conclusions} \label{SC}

In this paper we have applied a Lagrangian (antifield) BRST formalism
to the quantization of collective coordinates of a model describing the
motion of a particle on a Riemannian manifold under a scalar
potential.

To perform a semiclassical expansion for such a theory the minima of
the potential must be found, and the potential must be expanded around
them. This yields a set of harmonic oscilators plus higher order terms
which are treated in perturbation theory. As is well known, this scheme
fails if the potential has a manifold of minima. In such case there are
directions with zero restoring force which spoils the perturbation
theory with infrared divergences. The classical solution breaks a
symmetry which is quantically unbroken.

In order to restore the symmetry and to cure the infrared divergences
we introduced explicitly collective coordinates, as time dependent
parameters of the symmetry transformations of the action broken by the
classical solution. The group of symmetries of the action was
considered to be in general non-abelian.

The resulting gauge theory was quantized using the antifield
formalism. A gauge-fixed BRST invariant action was obtained. It was
also shown that the same result can be found by a hamiltonian BRST
procedure.

We took advantage of finite temperature techniques to calculate the
partition function. It was shown that a suitable choice of the
gauge-fixing functions decouples the fields at quadratic level and
provides non-zero spurious frequencies to the zero modes.

We calculated the intrinsic partition function at two-loops,
and the collective hamiltonian at one-loop. It was shown that these
results are independent of the gauge-fixing parameters, as expected.

The model considered in this paper encompasses, in the case of an
infinite number of coordinates, soliton models, as for example the
Skyrme model. However, the application of our results to such a system
would involve the numerical task of determining the normal modes of the
quadratic action, as well as the solution of ultraviolet problems. 

\acknowledgements

D.R.B.\ thanks Fundaci\'on Antorchas for an Antorchas Fellowship which
has partially sponsored this work. In addition, J.P.G.\ and
D.R.B.\ have been supported by the Consejo Nacional de Investigaciones
Cient\'{\i}ficas y T\'ecnicas (CONICET).

\appendix

\section*{}

When expanding the action in terms of powers of the normal modes
it is useful to introduce a number of coefficients whose definitions
we collect below:
\begin{eqnarray}
A_{nml}  &=& 
	\bar{V}_{stu} \psi^s_n \psi^t_m \psi^u_l ,\\
B_{nm;l} &=& 
	\bar{g}_{st,u} \psi^s_n \psi^t_m \psi^u_l ,\\
D^{a'}_{nm} &=& {\Im_{a'}}^{-\frac{1}{2}}
	\bar{g}_{st} \partial_{a'v}R^s(\bar{q},0) \psi^t_n \psi^v_m ,\\
D^{\bar{a}}_{nm} &=& 
	\bar{g}_{st} \partial_{\bar{a}v}R^s  \psi^t_n \psi^v_m ,\\
E_{nmlk} &=& 
	\bar{V}_{stuv} \psi_{n}^s \psi_{m}^t 
	\psi_{l}^u \psi_{k}^v ,\\
F_{nm;lk} &=& 
        \bar{g}_{st,uv} \psi^s_n \psi^t_m 
	\psi^u_l \psi^v_k ,\\
H^{a'}_{m;nl} &=& {\Im_{a'}}^{-\frac{1}{2}}
        g_{st} \partial_{auv}R^s(\bar{q},0) \psi^t_m \psi^u_n \psi^v_l . 
\end{eqnarray}
These coefficients are related to each other by the fact that the action
(\ref{S'}) is gauge invariant. To find all relations which are useful
to simplify the calculations we differentiate both sides of
Eqs.(\ref{varg}-\ref{varV}) with respect to $\alpha_a$ and evaluate
them at $\alpha_a=0$:
\begin{eqnarray}
\partial_s V[R(q,0)] \partial_a R^s &=& 0 , \\
\partial_a\partial_s R^u(q,0) g_{ut} + \partial_a\partial_t R^v(q,0) g_{sv} 
                        + \partial_aR^w(q,0) g_{st,w} &=& 0 ,
\end{eqnarray}
where we have used that $\partial_s R^t(q,0) = \delta_s^t$.

Expanding these equations around the classical minimum the desired relations
are obtained:
\begin{eqnarray}
E_{mnla'} &=&   - A_{kmn} D^{a'}_{kl} 
                - A_{kml} D^{a'}_{kn}
                - A_{knl} D^{a'}_{km}
	\nonumber \\ &&
                - {\tilde{\omega}}^2_m H^{a'}_{m;nl} 
                - {\tilde{\omega}}^2_n H^{a'}_{n;ml}  
                - {\tilde{\omega}}^2_l H^{a'}_{l;mn} , \\
A_{a'nm}   &=&  -{\tilde{\omega}}^2_n D^{a'}_{nm} 
                -{\tilde{\omega}}^2_m D^{a'}_{mn}    , \\
B_{nm;a'}  &=& -D^{a'}_{mn} - D^{a'}_{nm}            , \\
F_{nl;ma'} &=& -B_{kl;m} D^{a'}_{kn} - B_{nk;m} D^{a'}_{kl} 
               -B_{nl;k} D^{a'}_{km} 
	\nonumber \\ &&
	- H^{a'}_{l;nm} - H^{a'}_{n;ml} , \\ 
D^{\bar{a}}_{\bar{n}b'} &=&
	D^{\bar{a}}_{b'\bar{n}} = 0 \\
D^{\bar{a}}_{c'b'} &=& - \sqrt{\frac{\Im_{c'}}{\Im_{b'}}} 
	{C_{\bar{a}b'}}^{c'} , \\
{C_{\bar{a} c'}}^{b'} &=& - \frac{\Im_{c'}}{\Im_{b'}}
	{C_{\bar{a} b'}}^{c'} ,
\end{eqnarray}
where, for brevity, we define $\tilde{\omega}_{\bar{n}} =
\omega_{\bar{n}}$, $\tilde{\omega}_{a'}=0$. Further relations can be
obtained using the commutation relations (\ref{commR}):
\begin{eqnarray}
D^{a'}_{nb'} - D^{b'}_{na'} &=& 0 \\
D^{a'}_{mn} D^{b'}_{nl} - D^{b'}_{mn}D^{a'}_{nl} &=& 
      	\sqrt{\frac{\Im_{c'}}{\Im_{a'}\Im_{b'}}} 
	{C_{b'a'}}^{c'} D^{c'}_{ml}
     	\nonumber \\ &&
	+\sqrt{\frac{ 1 }{\Im_{a'}\Im_{b'}}} 
      	{C_{b'a'}}^{\bar{c}} D^{\bar{c}}_{ml} + 
	H^{b'}_{m;la'} - H^{a'}_{m;lb'} 
\end{eqnarray}

It is also useful to define some functions obtained integrating the
thermal propagator:
\begin{eqnarray}
g(\omega_n) &=& \int_0^{\beta} d\tau G_{\omega_n}(\tau) = 
	\frac{\beta}{\omega_n^2} \\
g(\omega_{n},\omega_{m},\omega_{l}) &=&
  	\int_0^{\beta} d\tau G_{\omega_n}(\tau)
	G_{\omega_m}(\tau)G_{\omega_l}(\tau) 
	\nonumber \\ &=& 
	\beta \left[ 
	\omega_n \left( \omega_n^2 - \omega_m^2 - \omega_l^2 \right) 
	\coth{\left( \frac{\beta \omega_m}{2} \right)} 
	\coth{\left( \frac{\beta \omega_l}{2} \right)} 
	\right. \nonumber \\ && +
	\omega_m \left( \omega_m^2 - \omega_n^2 - \omega_l^2 \right) 
	\coth{\left( \frac{\beta \omega_n}{2} \right)} 
	\coth{\left( \frac{\beta \omega_l}{2} \right)}  
	\nonumber \\ && \left. +
	\omega_l \left( \omega_l^2 - \omega_m^2 - \omega_n^2 \right) 
	\coth{\left( \frac{\beta \omega_m}{2} \right)} 
	\coth{\left( \frac{\beta \omega_n}{2} \right)} +   
	2 \omega_n \omega_m \omega_l \right] / 
	\nonumber \\ &&
	\left[4\omega_n\omega_m\omega_l
        \left( \omega_n + \omega_m + \omega_l \right) 
	\left( \omega_n + \omega_m - \omega_l \right) 
	\left( \omega_n - \omega_m + \omega_l \right)\right.
        \nonumber \\ && 
	\left.\left( \omega_n - \omega_m - \omega_l \right)\right] \\
g(\omega_{n},\dot{\omega}_{m},\dot{\omega}_{l}) &=&
  	\int_0^{\beta} d\tau G_{\omega_n}(\tau) 
	\partial_{\tau}G_{\omega_m}(\tau)
        \partial_{\tau}G_{\omega_l}(\tau)                       
	\nonumber \\ &=& 
	\beta \left[  
	\omega_n \left( \omega_n^2 - \omega_m^2 - \omega_l^2 \right) 
	+ \omega_m \left( \omega_m^2 - \omega_n^2 + \omega_l^2 \right) 
	\coth{\left( \frac{\beta \omega_m}{2} \right)} 
	\coth{\left( \frac{\beta \omega_n}{2} \right)}  
	\right. \nonumber \\ &&  +
	\omega_l \left( \omega_l^2 - \omega_n^2 + \omega_m^2 \right) 
	\coth{\left( \frac{\beta \omega_l}{2} \right)} 
	\coth{\left( \frac{\beta \omega_n}{2} \right)} +   
	\nonumber \\ && \left. +
	2 \omega_n \omega_m \omega_l 
	\coth{\left( \frac{\beta \omega_m}{2} \right)} 
	\coth{\left( \frac{\beta \omega_l}{2} \right)}  \right] /
	\nonumber \\ &&
	4\omega_n \left( \omega_n + \omega_m + \omega_l \right) 
	\left( \omega_n + \omega_m - \omega_l \right) 
	\left( \omega_n - \omega_m + \omega_l \right) 
	\left( \omega_n - \omega_m - \omega_l \right) \\
g(\omega_{n},\omega_{m}) &=& 
  	\int_0^{\beta} d\tau G_{\omega_n}(\tau)G_{\omega_m}(\tau) 
        \nonumber \\ &=&
	\frac{\beta}{2\omega_n\omega_m
         \left( \omega_n^2 - \omega_m^2 \right)}
	\left[ 
	\omega_n 
	\coth{\left( \frac{\beta \omega_m}{2} \right)} - 
	\omega_m 
	\coth{\left( \frac{\beta \omega_n}{2} \right)}
	\right] \\
g(\dot{\omega}_n,\dot{\omega}_m) &=&
  	\int_0^{\beta} d\tau \partial_{\tau}G_{\omega_n}(\tau) 
        \partial_{\tau}G_{\omega_m}(\tau)
	\nonumber \\ &=&
	\frac{\beta}{2\left( \omega_n^2 - \omega_m^2 \right)}
	\left[ 
	\omega_n 
	\coth{\left( \frac{\beta \omega_n}{2} \right)} - 
	\omega_m 
	\coth{\left( \frac{\beta \omega_m}{2} \right)}
	\right] 	            
\end{eqnarray}


\end{document}